\documentclass[12pt]{article}

\usepackage{graphicx}
\usepackage{amssymb}
\usepackage{MnSymbol}
\usepackage{amsmath}
\usepackage{lineno}

\begin{document}

\title{The classical geometrization of the electromagnetism}

\author{Celso de Araujo Duarte\footnote{E-mail: celso@fisica.ufpr.br} \\
\\Departamento de F\'isica, Universidade Federal do Paran\'a\\ CP 19044, 81531-990, Curitiba, PR, Brazil}

\maketitle

\begin{abstract}
Following the line of the history, if by one side the electromagnetic theory was consolidated on the 19th century, the emergence of the special and the general relativity theories on the 20th century opened possibilities of further developments, with the search for the unification of the gravitation and the electromagnetism on a single unified theory. Some attempts to the geometrization of the electromagnetism emerged in this context, where these first models resided strictly on a classical basis. Posteriorly, they were followed by more complete  and embracing quantum field theories. The present work reconsiders the classical viewpoint, with the purpose of showing that at first order of approximation the electromagnetism constitutes a geometric structure aside other phenomena as gravitation, and that magnetic monopoles do not exist at least up to this order of approximation. Even though being limited, the model is consistent and offers the possibility of an experimental test of validity.
\end{abstract}

\textit{Keywords}: electromagnetism; general relativity; Riemannian geometry; field equations; Lagrangian\\

DOI: 10.1142/S0219887815600221\\


\section{Introduction}

If by one side the roots of the electromagnetism stay on the qualitative description of a variety of phenomena (the electrification and the magnetization of bodies and their mutual attraction or repulsion), on the other side the contribution of 19th century researchers -- as Amp\`ere, Faraday, Biot and Savart -- left to the formulation of a quantitative basis. Their discoveries left to the inference that the electric and the magnetic phenomena obey to simple mathematical laws, which with the electromagnetic theory was constructed on a solid basis.

Subsequently, the evolution of the electromagnetic theory was characterized by an increasing mathematization, converging to the nowadays framework, which is based on the Maxwell equations \cite{Jackson}. However, such developments rested on algebra, but not on geometry.

Certainly, the solidity of the electromagnetic theory was a strong supporting basis when the negative results of the Michelson and Morley experiment arrived to the impasse around the existence of the aether, on the end of the 19th century \cite{Michelson,Whittaker}. The fall of the hypothesis of the aether with the contributions of Lorentz \cite{Whittaker,Lorentz}, 
Poincar\'e \cite{Whittaker2} and Einstein \cite{Einstein0} left finally to the emergence of the special relativity theory (SRT), with the fall down of the Newtonian conception of space and time as absolute quantities.

On the SRT, the invariability of space and time intervals $\Delta r$ and $\Delta t$ with respect to the change of the inertial reference frame was replaced by the invariability of the space-time interval $\Delta s$, which on the infinitesimal form is given as a function of the infinitesimal space and time intervals $dr=\sqrt{\left(dx\right)^2+\left(dy\right)^2+\left(dz\right)^2}$ and $dt$ by
\begin{equation}\label{space-time-SRT}
\left(ds\right)^2=c^2\left(dt\right)^2-\left(dx\right)^2-\left(dy\right)^2-\left(dz\right)^2
\end{equation}
where $c$ is the velocity of the light. The validity of such conception left to a correction on the Galilean transformations, with the emergence of the Lorentz transformations \cite{Lorentz}. An important outcome of the SRT on the terrain of the electromagnetism was the solving of an old issue, the lack of invariance of the Maxwell equations under the change of the reference frame under the Galilean transformations.

The next step was the development of the general relativity theory (GRT), which brought again the geometry to the scenario. Initially, Einstein and Grossmann generalized the expression \ref{space-time-SRT} to \cite{Whittaker2}
\begin{equation}\label{space-time-GRT}
ds^2=g_{\alpha\beta}dx^{\alpha}dx^{\beta}
\end{equation}
(using Einstein's summation notation for repeated indexes \cite{Einstein1}), where $g_{\alpha\beta}$ are the components of the symmetric, rank two metric tensor.

While the metric \ref{space-time-SRT} introduced the concept of a four-dimensional continuum with a diagonal Minkowskian metric $\eta_{\alpha\beta}$ with signature $(+---)$ \cite{Minkowski}, the generalization introduced by expression \ref{space-time-GRT} represented the break of the Euclidean character of the geometry of the space-time. On this new context, it is defined a non-zero space-time curvature defined by the components of the rank four Riemann curvature tensor $R_{\alpha\beta\mu\nu}$ \cite{Landau}. This Riemanian geometry was entirely determined by the ten components $g_{\alpha\beta}$ which played the role of gravitational potentials on the GRT \cite{Einstein1}.

Einstein, supported by Mach's principle \cite{Mach1,Mach2}, considered that the GRT was a generalization of the Newtonian law of gravitation. The rank two Ricci tensor $R_{\mu\nu}$ constructed by double-index contraction on the Riemann tensor, $R_{\mu\nu}=g^{\alpha\beta}R_{\alpha\mu\nu\beta}$ was employed for the construction of a classical field equation. This equation related the space-time curvature to the energy and matter densities represented by the stress-energy tensor $T_{\mu\nu}$. His considerations left him to the final expression \cite{Einstein1}
\begin{equation}\label{field-equation}
R_{\mu\nu}-\frac12g_{\mu\nu}R=\frac{8\pi G}{c^4}T_{\mu\nu}
\end{equation}
where $R=g^{\mu\nu}R_{\mu\nu}$ is the scalar curvature, $G$ is the Newton gravitational constant and $c$ is the velocity of light. Equations \ref{field-equation} are known as the Einstein field equations.

If by one side the GRT is the geometrization of the dynamics of the movement which explains the phenomenon of gravitation, it does not enframe the electromagnetism. Although, yet on the GRT there is an indirect participation of the electromagnetism via the energy-stress tensor, whose components depend on the electromagnetic energy and momentum densities. By the other side, we stress Einstein's statement that this tensor is only a ``weak spot of the theory'', since it is by itself devoid of any strict geometric meaning.

Posteriorly, very soon emerged new theories for the insertion of the electromagnetism on the geometry, especially on the period from 1914 to 1933 \cite{Goenner}. The first noticeable theory was that of Hermann Weyl, who introduced a group of one-dimensional gauge transformation on the metric \cite{Goenner}
\begin{equation}\label{Weyl-gauge}
g\rightarrow \bar{g}:=\lambda g
\end{equation}
Postulating a non metricity tensor $Q_{ijk}=Q_k g_{ij}$ associated to an arbitrary vector field $Q_k$, the transformation \ref{Weyl-gauge} implies that
\begin{equation}\label{non-metricity}
Q_k\rightarrow Q_k+\partial_k\sigma
\end{equation}
which is similar to the gauge transformation of the electromagnetic field. On the search for a scalar density Lagrangian, Weyl arrived to
\begin{equation}\label{Weil-action}
I=\int\sqrt{-g}\left(R^2+6F_{\mu\nu}F^{\mu\nu}\right\}
\end{equation}
where $F_{\mu\nu}$ and $F^{\mu\nu}$ are the covariant and the contravariant components of the electromagnetic field tensor \cite{Landau}.

Despite the Weyl theory was consistent with the Maxwell equations, a serious impasse was the issue of the measure of the length of vectors and their integrability. A length $l$ in the presence of the electromagnetic field would be given as a function of the corresponding zero field length $l_0$ by
\begin{equation}\label{l}
l=l_0 exp\left(A_{\mu}dx^{\mu}\right)
\end{equation}
being $A_{\mu}$ the components of the four vector electromagnetic potential. In this sense, a length would be affected by the flow of the time. As Einstein observed, this would reflect on the spacing of spectral emission lines that by this reason would depend on the history, declaring that:\\

\hangindent=20pt``\textit{Regrettably, the basic hypothesis of the theory seems unacceptable to me, [of a theory] the depth and audacity of which must fill every reader with admiration.''} \cite{Weyl}\\

As Goenner observes \cite{Goenner}, the gauging of fields was actually a central point of attention during the decade from 1918 to 1928. With the advent of the concept of spinor, Weyl tried to adapt his theory making an association between the electromagnetic field and the gauge for the wave function \cite{Weyl2}.

An important advance was achieved in 1921 with the new theory of Theodor Kaluza \cite{Kaluza}. His theory was based on the formalism of the GRT, adding a fourth space-like dimension $x_5$ to the space-time continuum, from which five additional components appeared on the metric, $g_{15}$,... $g_{55}$. The four components $g_{15}$,... $g_{45}$ were identified to the components of the four vector potential $A_1$,... $A_4$, and the fifth component $g_{55}$ was taken as a scalar gravitational potential. Kaluza considered a small deviation from the Minkowskian metric $\eta_{\mu\nu}$ by small quantities $\gamma_{\mu\nu}$,
\begin{equation}\label{g-Kaluza}
g_{\mu\nu}=\eta_{\mu\nu}+\gamma_{\mu\nu}
\end{equation}
from which he succeeded to establish a model for the inclusion of the electromagnetism on the space-time metric.

Soon after, Oskar Klein proposed on 1926 that Kaluza's fifth extra dimension would be compactified, curled up in a circle of a very small radius \cite{Klein}. In this sense, the motion of a particle along the fifth dimension would leave it again to its point of origin.

The basis of Kaluza’s theory was also the source of inspiration to subsequent models, as for example the more recent work of Edward Witten of 1981 \cite{Witten}.

On 1953, Wolfgang Pauli presented to Einstein, on a private correspondence, an six-dimensional formulation of the field equations \ref{field-equation}. However, there is no evidence that he had shown a field Lagrangian and its quantization. Declaring that ``[the model] \textit{leads to some rather unphysical shadow particles}'', he refrained to publish his idea \cite{Straumann}.

At this moment, it is relevant to stress the statement of Einstein, that ``\textit{in the end, things must arrange themselves such that action-densities need not be glued together additively}'' \cite{Schulman}. In this sense, no theory could be considered complete if this requirement is not satisfied -- which is the case of Weyl's action presented above (expression \ref{Weil-action}).

The demand of Einstein can be considered as conceptual requirement on the most refined degree. However, there is a priority demand: any complete and embracing theory should be consistent with the framework of the quantum theory. By this reason, the ``classical'' aspect of any theory should be adapted to fit the requirements of quantum mechanics, which is the purpose of the nowadays theories of quantum gravity, as the string theory and the loop quantum gravity.

Despite the fact that any geometric theory of the electromagnetism with strictly classical basis would not be embracing and complete, it may highlight key points that underlie the physical phenomena. Even if it does not fully embrace the multiple aspects of the problem, it may provide new insights for new theories.

In this spirit of ideas -- after this short and rough historical overview -- we frame the present work. Here it is achieved to a mathematical integration of the electromagnetism on the space-time metric under a classical framework, up to first order. Contrary to the common \textit{Ansatz}, the central point is not the reconstruction of abstract and underlying concepts, as the space-time interval $ds$, the metric $g_{\mu\nu}$, or the action of the field. In place of this, we appeal to a direct argumentation around the geodesic equation -- the equation of the movement of a charged particle in curved space-time. The starting point of motivation for the present work is an apparent redundancy on this equation. Additional motivating arguments are: 1. the representation of the Maxwell equations in similar way to the Einstein field equations; 2. the representation of the action density for the electromagnetism by a single term (no additive terms); 3. the explanation of the vanishing divergence of the magnetic field -- while not necessarily that of the electric field -- as a mathematical identity; 4. the conception that magnetic monopoles (if they exist) should be a second order effect (consequence of item 3); 5. the emergence of a ``charge-current tensor'' $Q_{\mu\nu}$ with similar nature to that of the energy-stress tensor; 6. the functional form of the space-like components $Q_{ab}$, resembling the three-dimensional shear-stress tensor (interpreted as the shear-stress of space). We stress that the present formalism, despite similar, has no relationship with the wellknown gravitoelectromagnetism.

Finally, it is suggested an experimental test for the present framework, based on the verification of the muonium lifetime.

This work is organized on the following way: after the mathematical basis presented on section \ref{Init}, section \ref{geoelectro} shows the apparent redundance on the geodesic equation, that leaves to a merge of the electric and the magnetic fields with the space-time metric. Section \ref{Efe} shows the correspondent Ricci curvature tensor, from which it is constructed a first order field equation for the electromagnetism. Then, it is introduced the concept of charge-current tensor, and considered the issue about the magnetic monopoles. Section \ref{lagrangian} presents an alternative first order Lagrangian of the electromagnetic field. Finally, section \ref{final} proposes an experimental test of validity of the present model and the final section presents the conclusions.

\section{Initial considerations}\label{Init}

On the GRT, the trajectory of a massive body with coordinates $(ct,x,y,x)=(x^0,x^1,x^2,x^3)$ in the presence of a gravitational field is described by the equation of the geodesic, which in affine curved space-time is given by \cite{Einstein1,Landau}
\begin{equation}\label{geodesic}
\frac{d^2x^{\mu}}{ds^2}+\Gamma^{\mu}_{\alpha\beta}\frac{dx^{\alpha}}{ds}\frac{dx^{\beta}}{ds}=0
\end{equation}
where $\Gamma^{\mu}_{\alpha\beta}$ are the Christoffel symbols. In this equation, the first term is the four-vector acceleration and the second represents the effect of gravity. Taking the approximation of small gravitational fields on equation \ref{geodesic}, the component $g_{00}$ of the metric tensor is identified to the Newtonian gravitational potential, and we arrive just to the equality of Newton's second law of the movement to the Newtonian gravitational force, owing to the principle of equivalence \cite{Einstein1,Landau}.

The presence of an electromagnetic field (EMF) is usually represented with the addition of a term to the second member of equation \ref{geodesic}, \cite{Landau}
\begin{equation}\label{geodesic2}
\frac{d^2x^{\mu}}{ds^2}+\Gamma^{\mu}_{\alpha\beta}\frac{dx^{\alpha}}{ds}\frac{dx^{\beta}}{ds}=\frac q{mc^2} F^{\mu}_{\ \ \nu}\frac{dx^{\nu}}{ds}
\end{equation}
where $F^{\mu}_{\ \ \nu}$ are the contravariant-covariant components of the electromagnetic field tensor, while $F_{\mu\nu}=A_{\mu,\nu}-A_{\nu,\mu}$ \cite{Landau}. On this equation, the effect of the curvature of the space-time enters  implicitly on the Christoffel, symbols while the electromagnetism enters on the EMF tensor.

Finally, with respect to the Maxwell equations, they can be written in terms of the electromagnetic tensor as
\begin{equation}\label{F}
\begin{array}{l}
F^{\mu\nu}_{\nu}=\mu_0j^{\mu}\\
F_{\alpha\beta,\gamma}+F_{\gamma\alpha,\beta}+F_{\beta\gamma,\alpha}+=0
\end{array}
\end{equation}
where $j^{\mu}$ is the four-vector current density. To account for the effect of the curvature of the space-time, the partial derivatives in this expression are replaced by covariant derivatives by the addition of terms with Christoffel symbols \cite{Landau}.

\section{From the geodesic equation to the electromagnetism}\label{geoelectro}

On the limit of small velocities ($v<<c$), all the spatial components of the four-velocity $dx^{\alpha}/ds$ are negligible with respect to the time component $dx^0/ds\approx1$. Consequeltly, the second term of \ref{geodesic} is reduced approximately to $\Gamma^{\mu}_{00}\left(\frac{dx^0}{ds}\right)^2\approx\Gamma^{\mu}_{00}$. On the other side, if we consider the first order corrections\footnote{We consider the symmetry of the Christoffel symbols under permutation of the two lower indexes. We use Roman letter indexes for summation restricted to the spatial coordinates; Greek letter indexes refer to space and time coordinates.} (linear in $dx^a/ds$), the correction term is $2\Gamma^{\mu}_{a0}\frac{dx^a}{ds}\frac{dx^0}{ds}\approx2\Gamma^{\mu}_{a0}\frac{dx^a}{ds}$. This development could be considered simply as a mathematical degree of approximation of the problem. However, here we point a striking similarity between the above terms with very well-known expressions, the Coulomb force $\vec{F}=q\vec E$ caused by an electric field $\vec E$ acting on a charge $q$, and the force $\vec F=q\vec v\times\vec B$ caused by a magnetic field $\vec B$ acting on a charge $q$ moving at a velocity $\vec v$ \cite{Jackson}. Since the former is independent on the velocity and the latter is linear on this magnitude, we wonder if the geodesic equation \ref{geodesic} contains itself both these expressions of electromagnetism and electrodynamics -- and in this sense equation \ref{geodesic2} would have a redundant term on the right side.

We immediately try to check if there is a correspondence between the components of electric ($\vec{E}$) and magnetic ($\vec{B}$) fields and the terms involving the Christoffel symbols\footnote{Note that the Christoffel symbols are not tensor magnitudes and $\vec E$ and $\vec B$ are vectors in the three-dimensional space and not four-dimensional magnitudes; they are actually components of the electromagnetic tensor $F^{\mu\nu}$. Consequently there is not \textit{a priori} any restriction to a relation between $\vec E$, $\vec B$ and the $\Gamma^{\mu}_{\alpha\beta}$.}. Since on the limit $v<<c$ the metric $g_{\alpha\beta}$ is approximatelly equal to the Minkowski metric $\eta_{\alpha\beta}=2(\delta_{\alpha0}-\frac12)\delta_{\alpha\beta}$, we find after some algebra

\begin{equation}\label{E}
E^a=\frac {mc^2}q\left(-g_{a0,0}+\frac{g_{00,a}}2\right)
\end{equation}
Since $\vec E=-\partial\vec A/\partial t-\nabla\phi$, the first and the second terms in the parenthesis in \ref{E} can be imediately identified with the time derivative of the vector potential $\vec A$ and to the gradient of the scalar electric potential $\phi$ (respectively). Then, for weak fields,
\begin{equation}\label{g}
\left\{
\begin{array}{l}
g_{00}=1-\dfrac{2q}{mc^2}\phi\\
g_{i0}=\dfrac{q}{mc} A_i
\end{array}
\right.
\end{equation}

To check the consistence, we treat the case of the magnetic field. Identifying $-2\Gamma^a_{b0}\frac{dx^b}{ds}$ to the components of the vector product $\vec v\times\vec B$ which appear on the Lorentz force (less to a factor $\frac qm$), we arrive to the following system of linear equations in the components of $\vec v$,
\begin{equation}\label{system}
\left(
\begin{array}{ccc}
\gamma_{11} & \gamma_{12}-\lambda_3 & \gamma_{13}+\lambda_2\\
\gamma_{21}+\lambda_3 & \gamma_{22} & \gamma_{23}-\lambda_1\\
\gamma_{31}-\lambda_2 & \gamma_{32}+\lambda_1 & \gamma_{33}\\
\end{array}
\right)
\left(
\begin{array}{c}
v^1\\
v^2\\
v^3
\end{array}
\right)=0
\end{equation}
where $\gamma_{ij}=\Gamma^i_{j0}$ and $\lambda_i=-\frac{cq}{2m}B^i$. The above linear system is satisfied for any arbitrary $\vec v$ if and only if the determinant of the 3$\times$3 square  matrix on \ref{system} is zero. Fortunately, this condition is satisfied identically with the choice of $\vec A$ as presented in equations \ref{g}, since $\vec B=\nabla\times\vec A$.

In this sense, the metric without gravitation in the presence of a weak EMF is
\begin{equation}\label{weakfield}
g=\left(
\begin{array}{rrrr}
1-2r\phi & r cA^1 & r cA^2 & r cA^3\\
r cA^1 & -1 & 0 & 0\\
r cA^2 & 0 & -1 & 0\\
r cA^3 & 0 & 0 & -1\\
\end{array}
\right)
\end{equation}
where $r=q/mc^2$. Note that this metric depends implicitly on the charge-to-mass ratio $q/m$, and consequently the equivalence principle cannot be applied on this context. The metric \ref{weakfield} remains to be dependent on properties of the test particle, reflecting that the space-time geometry experienced by this particle is not universal but its intrinsic property.

Note also that when $2r\phi<<1$,  $g_{00}\approx1$. This implies that for energies $q\phi<<mc^2$ (much lower than the rest mass energy of the test particle) and if $rcA^i\approx 0$ we recover the Minkowski metric (for instance, for an electron $2r\phi\approx3 MeV$).

We stress that spatial rotations on the above metric \ref{weakfield} preserve the usual vector transformations of the coordinates of the vector $\vec A$, a fact that serves as an additional proof of consistence for the integration of the four-vector potential $A^{\mu}=(\phi,c\vec A)$ in the metric. Which respect to the relativistic invariance the formalism is, as expected, invariant by Lorentz transformations in first order approximation, as can be directly verified.

Following this line of reasoning, the equation of movement of a charged particle in the presence of an EMF without gravitation can be expressed simply by equation \ref{geodesic} employing the metric \ref{weakfield}, and the usual second member in equation \ref{geodesic2} is unnecessary, \textit{i.e.} equation \ref{geodesic2} has a redundancy.

\section{The curvature of the space-time and the field equations for the EMF}\label{Efe}

It is interesting to study the impact of the the metric \ref{weakfield} on the Einstein field equations\footnote{For simplicity and without loss of generality, we present the equation without the contribution of the cosmological term $\Lambda g_{\alpha\beta}$.}. Explicitly, the components of the Ricci tensor are given by
\begin{equation}\label{Ricci}
R_{\alpha\beta}=\Gamma^{\mu}_{\alpha\beta,\mu}-\Gamma^{\mu}_{\alpha\mu,\beta}+\Gamma^{\mu}_{\alpha\beta}\Gamma^{\nu}_{\mu\nu}-\Gamma^{\mu}_{\alpha\nu}\Gamma^{\nu}_{\beta \mu}
\end{equation}
and the energy-stress tensor is
\begin{equation}\label{T}
T_{\alpha\beta}=
\left(
\begin{array}{cccc}
\rho & p_1 & p_2 & p_3\\
p_1 & \sigma_{11} & \sigma_{12} & \sigma_{13}\\
p_2 & \sigma_{21} & \sigma_{22} & \sigma_{23}\\
p_3 & \sigma_{32} & \sigma_{32} & \sigma_{33}\\
\end{array}
\right)
\end{equation}
where $\rho$ is the energy density per unit of volume, $p_i$ ($i=1,2,3$) are the components of the momentum density and $\sigma_{ij}$ ($i,j=1,2,3$) are components of  hydrostatic pressure and shear.

In our first order approximation, one of the components of the Ricci tensor is
\begin{equation}\label{00}
R_{00}=\nabla^2\left(\frac{g_ 0}2\right) - \frac1c\partial_ 0 (\nabla.\vec{g})
\end{equation}
where $g_0=g_{00}$ and $\vec{g}=(g_{01},g_{02},g_{03})$. Employing relations \ref{g} and from the equality of electrodynamics $\nabla.\vec A +c^{-2}\partial_0\phi=0$ (vanishing divergence of the four-vector electromagnetic potential \cite{Jackson}), we obtain
\begin{equation}
R_{00}=r\square\phi
\end{equation}
Since $R=-(8\pi G/c^4)T$ and $T_{00}=\rho$ ($\rho$ is the energy density) \cite{Landau}, in the absence of shear, $T=\rho$, and we arrive to the wave equation
\begin{equation}\label{waveeq}
\square\phi=-\frac{8\pi G}{rc^4}\rho
\end{equation}
But for the EMF in vacuum, $\rho=E^2/\epsilon_0$. Then,
\begin{equation}\label{electro}
\square\phi=-\frac{8\pi G}{rc^4}\frac1{\epsilon_0}\vec E^2
\end{equation}
However, the second member is has a second order term that must be neglected. Then we arrive to an ordinary wave equation of classical electrodynamics in the absence of charges, $\square\phi=0$. 

Having obtained a wave equation for $\phi$ from $R_{00}$, we now consider all the components of the Ricci curvature tensor at first order of approximation:
\begin{equation}\label{alternative2}
\begin{array}{l}
R_{\alpha\beta}=r
\left(
\begin{array}{cccc}
\square\phi & \frac c2\left(\nabla\times\vec B\right)_1 & \frac c2\left(\nabla\times\vec B\right)_2 & \frac c2\left(\nabla\times\vec B\right)_3 \\
\frac c2\left(\nabla\times\vec B\right)_1 & -E_{1,1} & -\frac 12\left(E_{1,2}+E_{2,1}\right) & -\frac 12\left(E_{1,3}+E_{3,1}\right) \\
\frac c2\left(\nabla\times\vec B\right)_2 & -\frac 12\left(E_{2,1}+E_{1,2}\right) & -E_{2,2} & -\frac 12\left(E_{2,3}+E_{3,2}\right) \\
\frac c2\left(\nabla\times\vec B\right)_3 & -\frac 12\left(E_{3,1}+E_{1,3}\right) & -\frac 12\left(E_{3,2}+E_{2,3}\right) & -E_{3,3} \\
\end{array}
\right)
\end{array}
\end{equation}
We wish to construct a field equation for the electromagnetism in similar form to the Einstein field equations \ref{field-equation}, representing the set of the Maxwell equations \ref{F},
\begin{equation}\label{maxwell0}
\left\{
\begin{array}{l}
\nabla\times\vec E=\dfrac{\partial\vec B}{\partial t}\\
\nabla\times\vec B=c^{-2}\dfrac{\partial\vec E}{\partial t}+\vec j\\
\nabla.\vec E=\dfrac{\rho_q}{\epsilon_0}\\
\nabla.\vec B=0\\
\end{array}
\right.
\end{equation}
where $\vec E$ and $\vec B$ are respectively the vectors electric field and magnetic field, and $\rho_q$ is the charge density.

The first equation of \ref{maxwell0} is yet implicit with the definition $\vec E=-\partial\vec A/\partial t-\nabla\phi$ and $\vec B=\nabla\times\vec A$. On the other side, from \ref{alternative2} we get the scalar curvature,
\begin{equation}\label{R}
R=g^{\alpha\beta}R_{\alpha\beta}=r\left(\square\phi-\nabla.\vec E\right)
\end{equation}
Considering the wave equation in the presence of sources (electric charge density) $\square\phi=\rho_q/\epsilon_0$, the third equation of \ref{maxwell0} is recovered imposing that $R=0$.

Note that the components $R_{0a}=R_{a0}$ are proportional to the components of $\nabla\times\vec B$, which is the left member of the second equation of \ref{maxwell0}. However, there is not a term with $\nabla\cdot\vec B$ in \ref{alternative2}. Curiously, the equation $\nabla\cdot\vec B=0$ emerges on a mathematical identity: starting from the components of the Riemann curvature tensor $R_{\alpha\beta\gamma\delta}$ \cite{Landau},
\begin{equation}\label{Riemann}
R_{\alpha\beta\gamma\delta}=g_{\alpha\nu}\left(\Gamma^{\nu}_{\beta\delta,\gamma}-\Gamma^{\nu}_{\beta\gamma,\delta}+\Gamma^{\mu}_{\beta\delta}\Gamma^{\nu}_{\gamma\mu}-\Gamma^{\mu}_{\beta\gamma}\Gamma^{\nu}_{\mu\delta}\right),
\end{equation}
we arrive to:
\begin{equation}\label{Riemann2}
\left\{
\begin{array}{l}
R_{0000}=R_{000a}=R_{00a0}=R_{0a00}=R_{a000}=R_{ab00}=R_{00ab}=R_{abcd}=0\\     
R_{a0c0}=-R_{a00c}=-R_{0ac0}=R_{0a0c}=-\frac r2\left(\frac{\partial E_a}{\partial x^c}+\frac{\partial E_c}{\partial x^a}\right)\\
R_{abc0}=-R_{ab0c}=R_{c0ab}=-R_{0cab}=\frac r2\frac{\partial}{\partial x^c}\left(\frac{\partial A^a}{\partial x^b}-\frac{\partial A^b}{\partial x^a}\right)
\end{array}
\right.
\end{equation}
From the first Bianchi identity \cite{Landau}, it follows that
\begin{equation}\label{Bianchi}
R_{1230}+R_{3120}+R_{2310}\equiv0
\end{equation}
On the other side, from \ref{Riemann2} we verify that the left hand side of \ref{Bianchi} is proportional to $\nabla\cdot\vec B$. Consequently, the fourth equation of \ref{maxwell0} is a mathematical identity, revealing that magnetic monopoles do not exist at first order approximation. Since our scale factors are $q\phi/mc^2$ and $q\vec A/mc$ (typically negligible) we infer that if magnetic monopoles exist, their densities $\rho_m$ should be huge to be detected as a second order effect.

After some algebra we arrive to the searched equation for the electromagnetism in the form:
\begin{equation}\label{electrofield}
R_{\alpha\beta}-\frac12g_{\alpha\beta}R=\lambda Q_{\alpha\beta}
\end{equation}
where\footnote{Note that $R=0$ and then the term $g_{\alpha\beta}R$ can be omitted.} $\lambda=\mu_0q/m$ and we defined the charge-current tensor $Q_{\alpha\beta}$,
\begin{equation}\label{Q}
\begin{array}{l}
Q_{\alpha\beta}=
\left(
\begin{array}{cccc}
\rho_q & -\frac 1{2c}k_1 & -\frac 1{2c}k_2 & -\frac 1{2c}k_3 \\
-\frac 1{2c}k_1 & s_{11} & s_{12} & s_{13} \\
-\frac 1{2c}k_2 & s_{21} & s_{22} & s_{23} \\
-\frac 1{2c}k_3 & s_{31} & s_{32} & s_{33} \\
\end{array}
\right)
\end{array}
\end{equation}
where we introduced the three-dimensional tensor
\begin{equation}\label{s}
s_{ij}=\frac{\epsilon_0}2\left(E_{i,j}+E_{j,i}\right),
\end{equation}
and the three-dimensional vector current $\vec k$ that comprises the current density $\vec j$ and the displacement current $\frac1c\partial_0\vec E$:
\begin{equation}
\vec k=\vec j+\frac1c\partial_0\vec E
\end{equation}

The right side of expression \ref{s} is formally similar to the strain tensor of the continuum mechanics linear elasticity theory \cite{Landau2}. In this way, the components $s_{ij}$ can be written in terms of a space-time strain tensor $\epsilon_{\alpha\beta}$,
\begin{equation}\label{s2}
s_{ij} = -\epsilon_{00,ij}+\left(\epsilon_{0i,j}+\epsilon_{ij,0}\right)_{,0}\\
\end{equation}
Since $E_i=-\partial_0A_i-\partial_0A_i$, we infer that $\epsilon_{00}=2A_0$ and $\epsilon_{0i}=-A_i$, which assign directly the components of the four vector potential to specific deformations of the three-dimensional space. With respect to the physical meaning, since the electric field is the areal density of electric flux $\Phi$, $E_i=d\Phi/dS_i$, the components $s_{ij}$ may be considered as the variations of the flux density along the directions of the space.

In summary, we interpret the above considerations as follows: if a charged test particle with mass $m$ and electric charge $q$ moves in an EMF, the effect of the Lorentz force is equivalent to the free motion of the particle around a geodesic in the space-time with the metric given by expression \ref{weakfield}. This metric depends on the charge and the mass of the particle, revealing the relativity of the geometry of space-time which is characteristic of each test particle. On the other side, the geometry of the space-time is determined by the electromagnetic field equations \ref{electrofield}, which relate the curvature and strain of space-time to electric charge and current densities. This completes the parallel between the formalism of the GRT and the electrodynamics.

\section{A generalized Lagrangian}\label{lagrangian}

If by one side the Lagrangian for the gravitational field on the GRT is given by 
\begin{equation}\label{L_G}
L_G=R=2g^{\alpha\beta}(\Gamma^{\gamma}_{\alpha[\beta,\gamma]}+\Gamma^{\mu}_{\nu[\mu}\Gamma^{\nu}_{\beta]\alpha}),
\end{equation}
which can be reduced to the simple form \cite{Landau}
\begin{equation}\label{L_G2}
L_G=2g^{\alpha\beta}\Gamma^{\mu}_{\nu[\mu}\Gamma^{\nu}_{\beta]\alpha},
\end{equation}
on the other side the Lagrangian for the electromagnetic field is
\begin{equation}\label{L_EM2}
L_{EM}=F^{\mu\nu}F_{\mu\nu}=g^{\mu\alpha}g^{\nu\beta}A_{[\alpha,\beta]}A_{[\mu,\nu]}
\end{equation}
The Lagrangians \ref{L_G} and \ref{L_EM2}  are bilinear on the first order derivatives of their corresponding potential fields $g_{\mu\nu}$ and $A_{\alpha}$. Consequently, both Lagrangians have similar functional structure.

As is well known, the Einstein field equations can be obtained variationaly from the Lagrangian \ref{L_G2} and the matter Lagrangian $L_M$ \cite{Landau}. On the other side, we pointed the similarity between \ref{electrofield} and the Einstein field equations \ref{field-equation}. Consequently, we conclude that a most general expression for the Lagrangian of the electromagnetic should be
\begin{equation}\label{L_EM3}
L_{EM}=R,
\end{equation}
In fact, after taking the variation of the action with a Lagrangian density $\sqrt{-g}R$, we obtain the left hand side of \ref{field-equation} which coincides, at first order approximation, with the right hand side of \ref{L_EM2}.


\section{Final consideration}\label{final}

An immediate consequence of our considerations would be the change of the lifetime of a charged particle in the presence of an external electromagnetic field. This could be observed on the muon ($\mu$), whose free lifetime $\tau_{\mu}$ is well known with high accuracy with a margin of error of around 30 ppm \cite{nakamura}. As a constituent of the muonium (Mu), for which the Bohr radius is $a_0=0.5$\AA, the average Coulomb potential energy of the muon in the presence of the electron field is $e\phi=-27.7 MeV$. Then, from the expression of the infinitesimal space-time interval $ds^2=g_{\mu\nu}dx^{\mu}dx^{\nu}$ and the metric $g_{\mu\nu}$ presented on \ref{g}, we deduce that the relation between the lifetimes of these free and bound muon is approximately
\begin{equation}\label{mu}
\tau_{\mu}=\left(1-\frac{e\phi}{m_{\mu}c^2}\right)\tau_{Mu}
\end{equation}
which represents a decrease of only 0.3 ppm and unfortunately is two orders of magnitude smaller than the present margin of error of $\tau_{\mu}$. Considering the possibility of a future achievement of higher precision values for this magnitude, an experimental verification of expression \ref{mu} would serve as a first test of validity of the model.

\section{Conclusions}\label{conclusions}

As a conclusion, it was shown a mathematical framework where the electromagnetism was directly related to the space-time geometry. The considerations pointed to the possibility of future tests of the model by the determination of the lifetime of electrically charged elementary particles on the presence of electromagnetic fields. If the present model is correct, two important cases would remain to be explored within this classical framework: the most general case, not restricted to the weak field regime, and the coexistence of electromagnetic and gravitational fields.

\section{Acknowledgements}
The author is very grateful to Prof. A. J. Accioly (CBPF) for the incentive.

\end{document}